\documentclass[12pt,russian,english,english]{article}
\usepackage[T1]{fontenc}
\usepackage[koi8-r,latin9]{inputenc}
\usepackage{esint}

\makeatletter

\DeclareRobustCommand{\cyrtext}{%
  \fontencoding{T2A}\selectfont\def\encodingdefault{T2A}}
\DeclareRobustCommand{\textcyr}[1]{\leavevmode{\cyrtext #1}}
\AtBeginDocument{\DeclareFontEncoding{T2A}{}{}}

\newtheorem{theorem}{Theorem}

\newtheorem{lemma}{Lemma}

\date{}

\usepackage{babel}

\makeatother

\usepackage{babel}
\begin{document}

\title{Analytic dynamics of one-dimensional particle system with strong
interaction}

\author{V. A. Malyshev}
\maketitle
\begin{abstract}
We study here the small time dynamics of $N$ electrons on the circle
with Coul;omb repulsive interaction and study the series for the velocities
(initially zero). The main result is the estimates of the convergence
radius from below. We explain how this result is related to the problem
of very fast establishing of direct electric current. 
\end{abstract}

\section{Introduction}

\paragraph{Model and results}

We consider a system of $N$ point particles $i=1,2,...,N$ on the
interval $[0,L)\in R$, with periodic boundary conditions, that is
on the circle $S_{L}$ of length $L$. Initially, they are situated
at the points
\[
0=x_{1}(0)<...<x_{N}(0)<L
\]
The trajectories $x_{i}(t)$ are defined by the following system of
$N$ equations 
\begin{equation}
\frac{d^{2}x_{i}}{dt^{2}}=-\frac{\partial U}{\partial x_{i}}+F(x_{i})\label{main_eq}
\end{equation}
where the interaction between the particles is

\[
U(\{x_{i}\})=\sum_{<i,i-1>}V(x_{i}-x_{i-1})
\]
where summations is over all pairs of nearest neighbors on the circle.
The Coulomb potential $V(x)=V(-x)=\frac{1}{r},r=|x|$ is assumed,
and we denote $f(r)=-\frac{dV(r)}{dr}=r^{-2}$ the interaction force.
Note that the potential is repulsive, infinite at zero and thus, the
particles cannot change their order . $F(x)$ is the external force.

In the papers \cite{fixed_int,fixed_per} we considered fixed points
of such systems. Dynamics is more complicated to understand. It is
standard that the solution of the system (\ref{main_eq}) exists on
all time interval and is unique (for any initial conditions), however
to get more detailed information about particle trajectories (for
sufficiently large $N$) is sufficiently difficult and demands elaboration
of special methods. If moreover $F$ is analytic, then it is well-known
\cite{Golubev}, that the solution can be presented as the convergent
power series in $t$ in some neighborhood of $t=0$. 

Here we consider natural initial conditions, that is for all $i$
\begin{equation}
\Delta=\Delta_{i}(0)=x_{i+1}(0)-x_{i}(0)=\frac{L}{N},v_{i}(0)=0\label{initial}
\end{equation}
 and moreover, it is convenient to put $x_{1}(0)=0$. Note that this
configuration is a fixed point for zero external force. We are looking
for the solution of the form
\begin{equation}
v_{i}(t)=\sum_{j=1}^{\infty}c_{i,j}t^{j},c_{i,j}=c_{i,j}^{(N)}\label{series_u}
\end{equation}
and get bounds for the convergence radius, dependent on $N$, of these
series under the initial conditions (\ref{initial}). 

\begin{theorem}

Let $F$ be analytic on the circle $S_{L}$. Then 
\begin{enumerate}
\item for $j=1,2,...$, there exist numbers $b_{j}<\infty$, not depending
on $N$ and such, that for all $i,j$ and any $N$
\[
|c_{ij}|<b_{j}N^{\frac{j-1}{2}}
\]

\item if moreover, for some constant $C_{F}>0$ and all $x$ and $k$ 
\[
|F^{(k)}(x)|\leq C_{F}^{k+1},
\]
 then there exists constant $0<\chi<\infty$, not depending on $N$
and such that for all $i,j$
\[
|c_{ij}|<\chi^{j}N^{\frac{5}{6}j-\frac{3}{2}}
\]
 
\end{enumerate}
\end{theorem}

It follows that the convergence radius $R=R(N)$ of the series (\ref{series_u})
has a bound from below $R>\chi^{-1}N^{-\frac{5}{6}}$. From the proof
of the first assertion of the theorem (see section 2.2) one can see
that the bound from above could be of the order $\frac{1}{\sqrt{N}}$,
but this not yet rigorously proved. During the proof of the second
assertion of the theorem we give the explicit bound for $\chi$. Also,
explicit formulas for $c_{ij},j=1,2,3,4$ are presented.

\paragraph{Why this model}

Mathematical problems of statistical physics are elaborated sufficiently
deeply for equilibrium systems on the lattice. But on the continuous
(euclidean) space only for particle systems with small inverse temperature
or small density. There are many other cases where the problems are
not even formulated on the mathematical level. One of such cases is
the direct electric current. It is described by Ohm's law on the macro
level. On the micro level, in any textbook on condensed matter physics,
it is described as a flow of free (or almost free) electrons, any
of which is accelerated by the external force and impeded by the external
media (crystal lattice). Physics and mathematics introduced a lot
of such one-particle models with constant accelerating force and various
models (about 20, historically the first such model is the famous
Drude's model of 1900) of external media, where the particles loose
their kinetic energy. 

The central question is where the accelerating force comes from. The
problem is that the power lines can have hundreds kilometers of length
but the external force acts only on the length of several meters of
the wire. In fact, there are even more problems with the direct electric
current. For example, one should explain why the speed of the flow
is permanent and sufficiently slow, but this regime is being established
almost immediately. We discussed the first problem in (\cite{why_1}),
but there was no Coulomb interaction there. We shall come back to
this problem in the next paper. Here we discuss the second problem.

Consider first a trivial case with constant $F\geq0$ and with initial
conditions (\ref{initial}). Then it is clear that for any $i$
\[
v_{i}(t)=Ft,x_{i}(t)=x_{i}(0)+\frac{Ft^{2}}{2}
\]
that is $x_{i}(t)$ are analytic in $t$ for any $t$. Moreover, to
reach the speed of order $1$ (for example $v_{i}(t)=1$) the time
of order $1$ is necessary. We shall see however that the situation
is completely different for non-constant $F$. Anticipating the events,
it is important to note that technically the difference occurs because
for constant $F$ all discrete derivatives in the recurrent equations
below are identically zero. 

If we could prove that the coefficients $c_{ij}$ in the expansion
$v_{i}=\sum_{j=1}^{\infty}c_{i,j}t^{j}$ grow as $N^{aj}$ for some
$a>0$, then the speed of order $1$ will be achieved much earlier
- for time of the order $t=N^{-a}$, that is almost immediately. We
cannot prove such result as it is but the estimates of the coefficients,
obtained here, make such result quite plausible. 

The techniques of the paper is apparently new. There are results for
small time dynamics of multi-particle systems (see, \cite{Lanford}-\cite{mal_dynamicalClusters})
but they are not applicable in our case because of very strong interaction.

\section{Proof}

\subsection{Equations for the coefficients}

Fix initial data $x_{i}(0),v_{i}(0)$ as in (\ref{initial}) and consider
the trajectories $x_{i}(t)\in S_{L}$ on the interval $0\leq t<t_{0}$
for some $t_{0}=t_{0}(N)>0$. Putting 
\[
\Delta_{i}(t)=x_{i+1}(t)-x_{i}(t),\Delta=\Delta_{i}(0)=\frac{L}{N}
\]
 we have the equations 
\[
\frac{dv_{i}}{dt}=f(x_{i}(t)-x_{i-1}(t))-f(x_{i+1}(t)-x_{i}(t))+F(x_{i}(t))
\]
 or
\[
\frac{dv_{i}}{dt}=f(\Delta+\int_{0}^{t}[v_{i}(t_{1})-v_{i-1}(t_{1})]dt_{1})-f(\Delta+\int_{0}^{t}[v_{i+1}(t_{1})-v_{i}(t_{1})])dt_{1})+
\]
 
\[
+F(x_{i}(0)+\int_{0}^{t}v_{i}(t_{1})dt_{1})
\]

\paragraph{Integral equations}

Equivalent system of integral equations is
\[
v_{i}(t)=\int_{0}^{t}[f(\Delta+\int_{0}^{t}[v_{i}(t_{1})-v_{i-1}(t_{1})]dt_{1})-
\]
 
\begin{equation}
-f(\Delta+\int_{0}^{t}[v_{i+1}(t_{1})-v_{i}(t_{1})]dt_{1})+F(x_{i}(0)+\int_{0}^{t}v_{i}(t_{1})dt_{1})]dt\label{intEqua}
\end{equation}
 and can be rewritten as follows
\begin{equation}
v_{i}(t)=\int_{0}^{t}((\Delta+R_{i-1}(t))^{-2}-(\Delta+R_{i}(t))^{-2}+F(x_{i}(0)+\int_{0}^{t}v_{i}(t_{1})dt_{1}))dt\label{intEq_f}
\end{equation}
 where 
\[
R_{i-1}(t)=\int_{0}^{t}(v_{i}(t_{1})-v_{i-1}(t_{1}))dt_{1}
\]

For the sequel we need some notation for discrete derivatives. Let
on the interval $[0,N]\subset Z$ with periodic boundary conditions
a function $g(i)$ be given (that is a periodic function on $Z$ with
period $N$). Let us call
\begin{equation}
(\nabla g)(i)=(\nabla^{+}g)(i)=g(i+1)-g(i),(\nabla^{-}g)(i)=g(i)-g(i-1)\label{product_diff}
\end{equation}
 its right and left derivative correspondingly. Note that they commute
and the Leibnitz formula holds 
\begin{equation}
\nabla^{+}(gf)(i)=f(i+1)(\nabla^{+}g)(i)+g(i)(\nabla^{+}f)(i)=(Sf)(\nabla^{+}g)+g(\nabla^{+}f)\label{productDiff}
\end{equation}
 where $S$ is the shift operator
\[
(Sf)(i)=f(i+1)
\]
Below discrete derivatives will act on the indices $i$. If the function
$f(i)$ does not depend on $i$, then its (discrete) differentiation
gives zero.

We come back to the main equations and rewrite them as follows
\[
v_{i}(t)=\int_{0}^{t}dt[(-\nabla^{-}((\Delta+R_{i}(t))^{-2})+F(x_{i}(0)+\int_{0}^{t}v_{i}(t_{1})dt_{1})]
\]
 The following representation of the integrand will be useful

\[
(\Delta+R_{i-1}(t))^{-2}-(\Delta+R_{i}(t))^{-2}+F(x_{i}(0)+\int_{0}^{t}v_{i}(t_{1})dt_{1})=
\]
\[
=\Delta^{-2}(1+\frac{R_{i-1}}{\Delta})^{-2}-\Delta^{-2}(1+\frac{R_{i}}{\Delta})^{-2}+F(x_{i}(0)+\int_{0}^{t}v_{i}(t_{1})dt_{1})=
\]
\[
=F(x_{i}(0))+\sum_{m=1}^{\infty}d_{m}\Delta^{-2-m}(R_{i-1}^{m}-R_{i}^{m})+[F(x_{i}(0)+\int_{0}^{t}v_{i}(t_{1})dt_{1})-F(x_{i}(0))]
\]
 where
\[
d_{m}=(-1)^{m}(m+1)
\]
If $F$ is analytic on $S_{L}$, then there exists $\epsilon>0$ sufficiently
small and such that for any $x_{0}\in S_{L}$ the following series
\[
F(x)=F(x_{0})+\sum_{k=1}^{\infty}\frac{F^{(k)}(x_{0})}{k!}(x-x_{0})^{k}
\]
converges for all $x\in[x_{0}-\epsilon,x_{0}+\epsilon)$. Then finally
\begin{equation}
v_{i}(t)=F(x_{i}(0))t+\int_{0}^{t}\sum_{m=1}^{\infty}d_{m}\Delta^{-2-m}[-\nabla^{-}R_{i}^{m})]dt+\sum_{k=1}^{\infty}\int_{0}^{t}\frac{F^{(k)}(x_{i}(0))(\int_{0}^{t}v_{i}(t_{1})dt_{1})^{k}}{k!}dt\label{Equat_v_i}
\end{equation}

\paragraph{Recurrent equations}

Using (\ref{series_u}) and
\begin{equation}
R_{i-1}(t)=\sum_{j=1}^{\infty}(c_{i,j}-c_{i-1,j})\frac{t^{j+1}}{j+1}\label{Eq_R_1}
\end{equation}
 
\begin{equation}
R_{i}-R_{i-1}=\sum_{j=1}^{\infty}(c_{i+1,j}-2c_{i,j}+c_{i-1,j})\frac{t^{j+1}}{j+1}\label{Eq_R_2}
\end{equation}
we see, substituting (\ref{series_u}) to (\ref{Equat_v_i}), that
the right-hand side of (\ref{Equat_v_i}) can also be presented as
a convergent series with well-defined coefficients.

We shall find $c_{i,j}$ by equating the coefficients of $t^{j}$.
For $j=1,2$ the equations give immediately 
\begin{equation}
c_{i1}=F(x_{i}(0)),c_{i,2}=0\label{eq_j_1}
\end{equation}
as other summands in the right side of (\ref{Equat_v_i}) have larger
order in $t$. For $j\geq3$ the equations for the coefficients of
$t^{j}$ are
\begin{equation}
c_{ij}=\frac{1}{j}[\sum_{m=1}^{\infty}d_{m}\Delta^{-2-m}(-\nabla^{-}R_{i}^{m})+\sum_{k=1}^{\infty}\frac{F^{(k)}(x_{i}(0))(\int_{0}^{t}v_{i}(t_{1})dt_{1})^{k}}{k!}]_{j-1}\label{coeff_1}
\end{equation}
where, for the power series $\phi(t)=\sum_{k=0}^{\infty}a_{k}t^{k}$,
we denote $[\phi(t)]_{j}=a_{j}$. For $j>2$ the coefficients $c_{i,j}$
can be found recursively, moreover $c_{ij}$ depend only on $c_{i,k}$
with $k\leq j-2$. In fact, the right-hand part of the equation for
$c_{i,j}$ does not contain $c_{i,k}$ with $k\geq j-1$, as due to
(\ref{Eq_R_1}), each of $c_{ik}$ appears together with $t^{k+1}$.

Then the main equations will be
\begin{equation}
c_{ij}=\frac{1}{j}\sum_{m=1}^{\infty}d_{m}\Delta^{-2-m}(-\nabla^{-}[(\sum_{j=1}^{\infty}(c_{i+1,j}-c_{i,j})\frac{t^{j+1}}{j+1})^{m}]_{j-1})+\label{coeff_2}
\end{equation}
 
\[
+\sum_{k=1}^{\infty}\frac{F^{(k)}(x_{i}(0))}{k!}[\sum_{j=1}^{\infty}c_{i,j}\frac{t^{j+1}}{j+1})^{k}]_{j-1}
\]
We have 
\begin{equation}
[(\sum_{j=1}^{\infty}c_{i,i}\frac{t^{j+1}}{j+1})^{k}]_{j-1}=\sum_{j_{1}+...+j_{m}=j-m-1}\frac{c_{i,j_{1}}}{j_{1}+1}...\frac{c_{i,j_{k}}}{j_{k}+1}\label{squareBrack_1}
\end{equation}
where $\sum_{j_{1}+...+j_{m}=j-m-1}$ is the sum over all ordered
arrays $j_{1},...,j_{k}$, such that
\begin{equation}
(j_{1}+1)+...+(j_{k}+1)=k+j_{1}+...+j_{k}=j-1\label{sum_j_i_1}
\end{equation}
 It follows
\begin{equation}
k\leq j_{1}+...+j_{k}=j-1-k\leq j-2,k\leq[\frac{j-1}{2}]\label{sum_j_i_2}
\end{equation}
Similarly
\[
[\sum_{j=1}^{\infty}(c_{i+1,j}-c_{i,j})\frac{t^{j+1}}{j+1})^{m}]_{j-1}=\sum_{j_{1},...,j_{m}}^{(j-1,m)}\frac{\nabla^{+}c_{i,j_{1}}}{j_{1}+1}...\frac{\nabla^{+}c_{i,j_{m}}}{j_{m}+1}
\]
and both (\ref{sum_j_i_1}) and (\ref{sum_j_i_2}) hold with $k$
instead of $m$. 

That is why the equations can be written as
\begin{equation}
c=Gc+c^{(0)}\label{operator_Equat}
\end{equation}
where $c$ is the vector $c=\{c_{ij}\}$, the free term $c^{(0)}=\{c_{ij}^{(0)}\}$
is simple 
\begin{equation}
c_{i1}^{(0)}=F(x_{i}(0)),c_{ij}^{(0)}=0,j\geq2\label{c_0_equat}
\end{equation}
and non-linear operator $G$ is defined by

\[
c_{i1}=c_{i1}^{(0)}=F(x_{i}(0)),c_{i2}=0
\]
 
\begin{equation}
c_{ij}=(Gc)_{ij}=-\sum_{m=1}^{[\frac{j-1}{2}]}\sum_{j_{1}+...+j_{m}=j-m-1}A_{ij}(m;j_{1},...,j_{m})+\sum_{k=1}^{[\frac{j-1}{2}]}\sum_{j_{1}+...+j_{k}=j-k-1}B_{ij}(k;j_{1},...,j_{k})\label{recurrent_equ}
\end{equation}
 for $j\geq3$, where 
\begin{equation}
A_{ij}(m;j_{1},...,j_{m})=\frac{1}{j}d_{m}\Delta^{-2-m}\nabla^{-}(\frac{\nabla^{+}c_{i,j_{1}}}{j_{1}+1}...\frac{\nabla^{+}c_{i,j_{m}}}{j_{m}+1})\label{A_equat_0}
\end{equation}
 
\begin{equation}
B_{ij}(k;j_{1},...,j_{k})=\frac{1}{j}\frac{1}{k!}F^{(k)}(x_{i}(0))\frac{c_{i,j_{1}}}{j_{1}+1}...\frac{c_{i,j_{k}}}{j_{k}+1}\label{B_equat_0}
\end{equation}
 Further on, $F_{i,k,q}$ will denote any discrete derivative $(\prod_{p=1}^{q}\nabla^{s(p)})F^{(k)}(x_{i}(0))$,
where $s(p)=\pm$. For the estimates the choice of $s(p)$ does not
matter. Put $F_{i,k}=F_{i,k,0}$.

Explicit expression for $c_{i3},c_{i4}$ is immediately obtained if
in the equations (\ref{recurrent_equ}) we take into account only
the terms with $k=1$ and $m=1$, as $k.m\leq[\frac{j-1}{2}]\leq1$)

\[
c_{i3}=-\frac{1}{3}d_{1}\Delta^{-3}\nabla^{-}\nabla^{+}\frac{c_{i1}}{2}+\frac{1}{3}F^{(1)}(x_{i})\frac{c_{i1}}{2}=\frac{1}{6}(d_{1}\Delta^{-3}F_{i,0,2}+F_{i,0,0}F_{i,1,0})
\]
 
\[
c_{i4}=-\frac{1}{4}d_{1}\Delta^{-3}\nabla^{-}(\nabla^{+}\frac{c_{i1}}{2})^{2}+\frac{1}{4}F^{(1)}(x_{i})\frac{c_{i1}^{2}}{4}=
\]
 
\[
=\frac{1}{8}(-d_{1}\Delta^{-3}F_{i,0,2}F_{i,0;1}+\frac{1}{2}F_{i,1,0}F_{i,0,0}^{2})
\]
 From the formulas 
\begin{equation}
F_{i,0,1}=F_{i+1,0,0}-F_{i,0,0}=\int_{x_{i}}^{x_{i+1}}F^{(1)}(x)dx,|F_{i,0,1}|\leq C_{F}^{2}\Delta\label{deriv_bound}
\end{equation}
 
\[
F_{i,0,2}=(F_{i+2,0,0}-F_{i+1,0,0})-(F_{i+1,0,0}-F_{i,0,0})=\int_{x_{i}}^{x_{i+1}}(\int_{x}^{x+\Delta}F^{(2)}(y)dy)dx,|F_{i,0,2}|\leq C_{F}^{3}\Delta^{2}
\]
 it follows that
\[
|c_{i3}|\leq\frac{1}{3}C_{F}^{3}(\Delta^{-1}+\frac{1}{2}),|c_{i4}|\leq\frac{1}{4}C_{F}^{5}+\frac{1}{16}C_{F}^{4}
\]
It is easy to see that for most $i$ the coefficients $c_{i,3}$ are
really of the order $\Delta^{-1}$.

\subsection{Principal exponent}

From the recurrent formulas (\ref{A_equat_0}) and (\ref{B_equat_0})
it follows that $c_{ij}$ are finite and depend on $i,j,N$. First
of all, we shall study them as functions of $N$ for fixed $i,j$.
Otherwise speaking, we shall prove the first part of the theorem.
Define the principal exponent
\[
I(\xi)=\limsup_{N\to\infty}\frac{\ln|\xi|}{\ln N}
\]
for a variable $\xi$ depending on $N$. Roughly speaking, it shows
that the main order of the asymptotics of $\xi$ is $N^{I(\xi)}$.

We shall consider the algebra $\mathbf{A}$ of polynomials of countable
number of (commuting) variables $F_{i,k,q},i=1,...,N;k,q=0,1,2,...$
with real coefficients, not depending on $F$. For any monomial $M$
of this algebra denote 
\[
Q(M)=-\sum q
\]
over all $q$ in this monomial. The natural mapping of $\mathbf{A}$
onto its subalgebra $\mathbf{A}_{0}$, generated by all $F_{i,k}=F_{i,k,0}$,
is defined by the subsequent substitutions
\[
F_{i,k,q}=F_{i+1,k,q-1}-F_{i,k,q-1}
\]
 or with the following formula
\[
(\nabla^{+})^{n}=(S-1)^{n}=\sum_{k=0}^{n}C_{n}^{k}(-1)^{k}S^{n-k}
\]

\begin{lemma}\label{lemma_I_Q} For any monomial $M\in\mathbf{A}$

\[
I(M)\leq Q(M)
\]
 \end{lemma}

It is sufficient to prove that
\[
I(F_{i,q})\leq Q(F_{i,q})=-q
\]
 This can be done as in (\ref{deriv_bound}), using induction in $q$.
Put\inputencoding{koi8-r}\foreignlanguage{russian}{
\[
g_{i,n}=\nabla^{n}g_{i},\Delta=\frac{1}{N}
\]
Then
\[
g_{i,n+1}=\nabla^{n+1}g_{i}=g_{i+1,n}-g_{i,n}=\int_{x_{i}}^{x_{i}+\Delta}dy_{1}(\int_{y_{1}}^{y_{1}+\Delta}dy_{2}....(\int_{y_{n-1}}^{y_{n-1}+\Delta}dy_{n}g^{(n)}(y_{n})))
\]
and if $|g^{(n)}(x)|\leq C_{g}^{n+1}$, we have 
\[
|g_{i,n}|\leq C_{g}^{n+1}\Delta^{n}
\]
}

\inputencoding{latin9}For any polynomial $P=\sum a_{r}M_{r}$ with
(different) monomials $M_{r}$ and coefficients $a_{r}$, not depending
on $F$, but possibly depending on $N$, define 
\[
Q(P)=\max_{r}(I(a_{r})+Q(M_{r}))
\]
in agreement with the previous definition. Then for any polynomial
$P$
\[
I(P)\leq\max_{r}(I(a_{r})+I(M_{r}))\leq\max_{r}(I(a_{r})+Q(M_{r}))
\]
Note that for any two polyonomials 
\[
Q(P_{1}P_{2})\leq Q(P_{1})+Q(P_{2})
\]
 Also
\begin{equation}
Q(\nabla^{+}P)\leq Q(P)-1,Q(\nabla^{-}\nabla^{+}P)\leq Q(P)-2\label{Q_1_2}
\end{equation}
 Denote $\deg P$ the degree of the polynomial $P=\sum a_{r}M_{r}$,
that is the maximal degree of its monomials.

\begin{lemma}\label{lemma_degree}

For $j>2$ $c_{ij}$ is a polynomial in the algebra $\mathbf{A}_{0}$
and has degree not greater than $j-1$.

\end{lemma}

We already saw this for $j=3,4$. Note that
\[
\deg(\nabla^{\pm}P)=\deg P
\]
 Now one can use induction: in the formula (\ref{A_equat_0}) the
degree is $j-2$, and in (\ref{B_equat_0}) the degree will be $j-1$.

Note that the recurrent formulas define $c_{ij}$ for all functions
$F(x)$, not necessary analytic. That is why the following assertion
makes sense.

\begin{lemma}\label{lemma_bound} Let $F$ be infinitely differentiable.
Then for all $i,j$ 
\[
I(c_{ij})\leq Q(c_{ij})\leq\frac{j-1}{2}
\]

\end{lemma}

As
\[
Q(c_{ij})=0,j=1,2,Q(c_{i,3})=1,Q(c_{i,4})=0
\]
 then the assertion holds for $j=1,2,3,4$. We shall prove the lemma
by induction in $j$. Assume that
\[
Q(c_{ij})\leq\frac{j-1}{2}
\]
 for all $j=1,2,...,J-2$.

Then for given $m,j_{1},...,j_{m}$, accordingly to (\ref{sum_j_i_1})
and (\ref{sum_j_i_2}), we have
\[
Q(A_{iJ}(m;j_{1},...,j_{m}))\leq2+m-1+Q(c_{ij_{1}})+...+Q(c_{ij_{m}})-m\leq1+\frac{1}{2}(j_{1}+...+j_{m})-\frac{m}{2}=
\]
 
\[
=1+\frac{1}{2}(J-m-1)-\frac{m}{2}
\]
as, according to (\ref{Q_1_2}), $(-1)$ and $(-m)$ are appended
because of the discrete differentiation of the corresponding monomials.
The last expression attains its maximum when $m=1$. It follows
\[
Q(A_{iJ}(m;j_{1},...,j_{m}))\leq\frac{J-1}{2}
\]
Similarly, for $B_{iJ}(k:j_{1},...,j_{k})$ we have the following
inequalities
\[
Q(B_{iJ}(k;j_{1},...,j_{k}))\leq\frac{1}{2}(J-1-k)-\frac{k}{2}<\frac{J-1}{2}
\]
We get thus $Q(c_{iJ})\leq\frac{J-1}{2}$, and $I(c_{iJ})\leq Q(c_{iJ})\leq\frac{J-1}{2}$.

\subsection{Convergence radius}

Here we shall prove the second assertion of the theorem 1. In the
proof it is convenient to write $N$ instead of $\frac{N}{L}$ and
assume that $C_{F}\geq1$.

We shall use the following majorization principle for infinite system
of recurrent equations and inequalities: for example, if two systems
of equations are given
\[
c_{ij}^{(q)}=P^{(q)}(c_{i1}^{(q)},...,c_{i,j-2}^{(q)}),q=1,2
\]
where $P^{(q)}$ are the polynomials with coefficients $p_{\alpha}^{(q)}$,
where $p_{\alpha}^{(2)}\geq0,|p_{\alpha}^{(1)}|\leq p_{\alpha}^{(2)}$
for all $\alpha$, and also $|c_{ij}^{(1)}|\leq c_{ij}^{(2)}$ for
$j=1,2,3,4$, then $|c_{ij}^{(1)}|\leq c_{ij}^{(2)}$ for all $j$.
One of such system (we use it in the proof) corresponds to one-particle
problem (that is with $N=1$) with specially chosen external force,
will be now introduced. Other auxiliary system $\beta(c_{ij})$ with
positive coefficients will be introduced later.

\paragraph{One-particle problem}

For $j=1,2,...$ and fixed $a$ put 
\[
g_{j}=g_{j}(\frac{a}{2})=\{\frac{a}{2}\}^{j}\frac{1.3...(2j-1)}{j!}=\{\frac{a}{2}\}^{j}\frac{(2j)!}{2^{j}j!j!}\sim\{\frac{a}{2}\}^{j}\frac{1}{\sqrt{4\pi j}}
\]
 Then we have

\begin{lemma}\label{lemma_one_particle}

For $j=5,6,...$ the following inequalities hold
\[
g_{j}\geq\frac{1}{j}\sum_{k=1}^{[\frac{j-1}{2}]}\sum_{j_{1}+...+j_{m}=j-m-1}(\frac{a}{2})^{k+1}\frac{(k+1)(k+2)}{2}\frac{g_{j_{1}}}{j_{1}+1}...\frac{g_{j_{k}}}{j_{k}+1}
\]

\end{lemma}

Proof. Let the particle, situated initially at the point $x(0)=0$,
move with the speed ($a>0$ being arbitrary)
\[
v(t)=\frac{1}{\sqrt{1-at}}=\sum_{j=0}^{\infty}g_{j}t^{j}
\]
 in the field of the external force $F(x)$, which we should find.
Then
\[
x(t)=\int_{0}^{t}v(s)ds=(-\frac{2}{a})\sqrt{1-at}+\frac{2}{a}\Longrightarrow1-at=(1-\frac{ax}{2})^{2}
\]
 
\[
F=\frac{dv}{dt}=\frac{a}{2}\frac{1}{(1-at)^{\frac{3}{2}}}=\frac{a}{2}\frac{1}{(1-\frac{ax}{2})^{3}}
\]
 
\[
\frac{F^{(k)}}{k!}=(\frac{a}{2})^{k+1}\frac{3.4...(k+2)}{k!}=(\frac{a}{2})^{k+1}\frac{(k+1)(k+2)}{2}
\]
Similarly to the above derivation of the recurrent equations (the
only difference is that here $v(0)=1$ and there are no $A$-terms),
we get

\[
g_{j}=\sum_{k=1}^{[\frac{j-1}{2}]}\sum_{p=0}^{k-1}\sum_{j_{1}+...+j_{m}=j-m-1}\frac{1}{j}\frac{1}{k!}F^{(k)}(x_{i}(0))C_{k}^{p}v^{p}(0)\frac{g_{j_{1}}}{j_{1}+1}...\frac{g_{j_{k-p}}}{j_{k-p}+1}
\]
for $g_{j}$ defined above. Here, as above, $\sum_{j_{1}+...+j_{m}=j-m-1}$
is the summation over all $j_{1},...,j_{k-p}$ such that
\[
j_{1}+...+j_{k-p}=j-k-1
\]
As all coefficients are positive, then, neglecting the terms with
$p>0$, we have for any $a>0$
\[
\frac{1}{j}\sum_{k=1}^{[\frac{j-1}{2}]}\sum_{j_{1}+...+j_{m}=j-m-1}(\frac{a}{2})^{k+1}\frac{g_{j_{1}}}{j_{1}+1}...\frac{g_{j_{k}}}{j_{k}+1}\leq
\]
 
\[
\leq\frac{1}{j}\sum_{k=1}^{[\frac{j-1}{2}]}\sum_{j_{1}+...+j_{m}=j-m-1}(\frac{a}{2})^{k+1}\frac{(k+1)(k+2)}{2}\frac{g_{j_{1}}}{j_{1}+1}...\frac{g_{j_{k}}}{j_{k}+1}\leq g_{j}
\]

\paragraph{Majorization }

From the recurrent formula for $c_{ij}$ one can see that they can
be written as
\[
c_{ij}=\sum_{r=1}^{d_{ij}}b_{i,j,r}N^{I_{i,j,r}}M_{i,j,r}
\]
where $b_{i,j,r}$ amd $d_{ij}$ are some numbers not depending neither
on $N$ nor on $F$, and $M_{i,j,r}\in\mathbf{A}$. By lemma \ref{lemma_degree}
\[
\deg M_{i,j,r}\leq j-1
\]

We need other preliminary definitions. For any polynomial
\[
P=\sum b_{r}N^{I_{r}}M_{r}
\]
where $b_{r}$ are the numbers, not depending neither on $N$ nor
on $F$, $M_{r}\in\mathbf{A}$, we put 
\[
\beta(P)=\sum_{r}|b_{r}|N^{I_{r}+Q(M_{r})}C_{F}^{Q_{0}(M_{r})-Q(M_{r})+\deg M_{r}}
\]
where, for any monomial $M_{r}$, the natural number $Q_{0}(M_{r})$
equals the sum $\sum k$ over all factors (variables) $F_{i,k,q}$.
In particular,
\[
\beta(c_{ij})=\sum_{r}|b_{i,j,r}|N^{I_{i,j,r}+Q(M_{i,j,r})}C_{F}^{Q_{0}(M_{i,j,r})-Q(M_{i,j,r})+\deg M_{i,j,r}}
\]
 By definition for any $i,k$ 
\[
\beta(F_{i,k,0})=C_{F}^{k+1},\beta(\nabla^{\pm}F_{i,k,0})=\beta(F_{i,k,1})=C_{F}^{k+2}N^{-1}=C_{F}N^{-1}\beta(F_{i,k})
\]
 Moreover, for any two polynomials $P_{1},P_{2}$
\begin{equation}
\beta(P_{1}+P_{2})\leq\beta(P_{1})+\beta(P_{2}),\beta(P_{1}P_{2})\leq\beta(P_{1})\beta(P_{2})\label{subadditivity}
\end{equation}
 and for any monomial $M$
\begin{equation}
\beta(\nabla^{\pm}M)\leq(\deg M)N^{Q(M)-1}C_{F}^{Q_{0}(M)-Q(M)+\deg M+1}=(\deg M)C_{F}N^{-1}\beta(M)\label{beta_deriv_monome}
\end{equation}
 It follows that for any polynomial $P$
\begin{equation}
\beta(\nabla^{\pm}P)\leq(\deg P)C_{F}N^{-1}\beta(P)\label{beta_deriv_polinom}
\end{equation}
 We call $\beta(P)$ the majorant of $P$ as by
\[
|F_{i,k,1}|=|\nabla^{+}F_{i,k}|\leq\int_{x_{i}}^{x_{i+1}}|F_{i,k+1}(x)|dx\leq C_{F}^{k+2}N^{-1}=\beta(F_{i,k,1})
\]
 we have 
\[
|P|\leq\beta(P)
\]
 From (\ref{subadditivity}) and (\ref{operator_Equat}) it follows
that
\[
\beta(c_{ij})\leq\sum_{m=1}^{[\frac{j-1}{2}]}\sum_{j_{1}+...+j_{m}=j-m-1}\beta(A_{ij}(m;j_{1},...,j_{m}))+\sum_{k=1}^{[\frac{j-1}{2}]}\sum_{j_{1}+...+j_{k}=j-k-1}\beta(B_{ij}(k;j_{1},...,j_{k}))
\]

Our inductive hypothesis will be (\textcyr{\char241} $g_{j}=g_{j}(1)$)
\begin{equation}
\beta(c_{ij})\leq\chi^{j}N^{\frac{5}{6}j-\frac{3}{2}}g_{j},j=1,2,...,J-2\label{inductive}
\end{equation}

\paragraph{Initial data}

One can choose $\chi_{0}>0$ so that for $j-1,2,3,4$ 
\[
\chi_{0}^{j}N^{\frac{5}{6}J-\frac{3}{2}}g_{j}\geq\beta(c_{ij})
\]
In fact, only for $j=3$ there is dependence on $N$, but $\frac{5}{6}3-\frac{3}{2}$
is exactly $N$.

\paragraph{Inductive step for $A$-terms with $m>1$}

To estimate $A$-terms we distinguish two cases: $m=1$ and $m>1$.
For $m>1$ we use obvious bounds
\[
\beta(\nabla^{\pm}c_{ij})\leq2\beta(c_{ij}),\beta(\nabla^{-}(\nabla^{+}c_{i,j_{1}}...\nabla^{+}c_{i,j_{m}}))\leq2^{m+1}\prod_{p}\beta(c_{i,j_{p}})
\]
 Then from (\ref{subadditivity}) and (\ref{A_equat_0}) we get

\[
\beta(A_{iJ}(m;j_{1},...,j_{m}))\leq\frac{m+1}{J}N^{2+m}2^{m+1}\frac{\beta(c_{i,j_{1}})}{j_{1}+1}...\frac{\beta(c_{i,j_{m}})}{j_{m}+1}\leq
\]
 
\[
\leq\frac{m+1}{J}N^{2+m}2^{m+1}\chi^{J-m-1}N^{\frac{5}{6}(J-m-1)-m\frac{3}{2}}\prod_{p=1}^{m}\frac{g_{j_{p}}}{j_{p}+1}\leq
\]
 
\[
\leq2^{m+1}\chi^{J-m-1}N^{\frac{5}{6}J-\frac{3}{2}}\frac{m+1}{J}(\prod_{p=1}^{m}\frac{g_{j_{p}}}{j_{p}+1})
\]
as the exponent over $N$ for $m\geq2$ has the bound
\[
2+m+\frac{5}{6}(J-m-1)-m\frac{3}{2}=\frac{5}{6}J-m\frac{8}{6}+\frac{7}{6}\leq\frac{5}{6}J-\frac{3}{2}
\]
 Then by lemma \ref{lemma_one_particle} with $a=2$ (if $\chi\geq2$)
\[
\sum_{m=2}^{[\frac{j-1}{2}]}\sum_{j_{1}+...+j_{m}=j-m-1}\beta(A_{iJ}(m;j_{1},...,j_{m}))\leq
\]
 
\[
\leq N^{\frac{5}{6}J-\frac{3}{2}}\chi^{J}\sum_{m=2}^{\frac{j-1}{2}}2^{m+1}\chi^{-m-1}\sum_{j_{1}+...+j_{m}=j-m-1}\frac{m+1}{J}(\prod_{p=1}^{m}\frac{g_{j_{p}}}{j_{p}+1})\leq
\]
 
\begin{equation}
\leq(\frac{2}{\chi})^{3}N^{\frac{5}{6}J-\frac{3}{2}}\chi^{J}g_{j}\label{final_A}
\end{equation}

\paragraph{Inductive step for $A$-terms with $m=1$}

In case $m=1$ we shall use the following bounds for $j=J-2\geq3$.
By (\ref{beta_deriv_polinom}), (\ref{inductive}) and lemma \ref{lemma_degree}
\[
\beta(\nabla^{+}c_{ij})\leq(j-1))N^{-1}C_{F}\beta(c_{ij})\leq(j-1))N^{-1}C_{F}\chi^{j}N^{\frac{5}{6}j-\frac{3}{2}}g_{j}
\]
 Similarly
\[
\beta(\nabla^{-}\nabla^{+}c_{i,j})|\leq((j-1)C_{F})^{2}N^{-2}\chi^{j}N^{\frac{5}{6}j-\frac{3}{2}}g_{j}
\]
 This gives additional summand $(-2)$ in the exponent over $N$,
equal to
\[
3-2+\frac{5}{6}(J-2)-\frac{3}{2}\leq\frac{5}{6}J-\frac{3}{2}
\]
that is 
\[
A_{ij}(1;j_{1})=A_{ij}(1;J-2)=\frac{1}{j}|d_{1}|N^{3}\frac{\nabla^{-}\nabla^{+}c_{i,J-2}}{J-1}\leq
\]
 
\begin{equation}
\leq2C_{F}^{2}\chi^{j-2}N^{\frac{5}{6}J-\frac{3}{2}}g_{j-2}\leq\frac{2C_{F}^{2}g_{j-2}}{\chi^{2}g_{j}}\chi^{j}N^{\frac{5}{6}J-\frac{3}{2}}g_{j}\label{final_A_1}
\end{equation}

\paragraph{Inductive step for $B$-terms}

For $B$-terms the inductive bound is easier, but here the degree
of monomials increases
\[
\beta(B_{ij}(k;j_{1},...,j_{k}))=\frac{1}{j}\frac{1}{k!}\beta(F^{(k)}(x_{i}(0))\frac{c_{i,j_{1}}}{j_{1}+1}...\frac{c_{i,j_{k}}}{j_{k}+1})\leq
\]
 
\[
\leq\frac{1}{j}\frac{C_{F}^{k+1}}{k!}\frac{1}{j_{1}+1}...\frac{1}{j_{k}+1}\beta(c_{i,j_{1}})...\beta(c_{i,j_{k}})\leq\frac{1}{j}\frac{C_{F}^{k+1}}{k!}\frac{g_{j_{1}}}{j_{1}+1}...\frac{g_{j_{k}}}{j_{k}+1}\chi^{j_{1}+---+j_{k}}N^{j_{1}+---+j_{k}}
\]
 By lemma \ref{lemma_one_particle} 
\[
\sum_{k=1}^{[\frac{j-1}{2}]}\sum_{j_{1}+...+j_{k}=j-k-1}\beta(B_{ij}(k;j_{1},...,j_{k}))\leq\frac{1}{j}\sum_{k=1}^{[\frac{j-1}{2}]}\frac{C_{F}^{k+1}}{k!}\sum_{j_{1}+...+j_{k}=j-k-1}\frac{g_{j_{1}}}{j_{1}+1}...\frac{g_{j_{k}}}{j_{k}+1}\chi^{J-k-1}N^{\frac{5}{7}j}\leq
\]
 
\begin{equation}
\leq C_{F}e^{C_{F}}\chi^{J-2}N^{\frac{J}{2}}\frac{1}{j}\sum_{k=1}^{[\frac{j-1}{2}]}\sum_{j_{1}+...+j_{k}=j-k-1}\frac{g_{j_{1}}}{j_{1}+1}...\frac{g_{j_{k}}}{j_{k}+1}\leq C_{F}e^{C_{F}}\chi^{J-2}N^{\frac{J}{2}}g_{j}\label{final_B}
\end{equation}

Finally, to end the proof of the theorem, we sum up three obtained
expression (\ref{final_A}),(\ref{final_A_1}),(\ref{final_B}), and
choose $\chi=\chi_{1}>0$ so that 
\[
(8\chi^{-1}+\frac{2C_{F}^{2}g_{j-2}}{g_{j}}+C_{F}e^{C_{F}})\chi^{-2}\leq1
\]
 Then for any $\chi\geq\max(\chi_{0},\chi_{1})$ we will have
\[
|c_{ij}|\leq\chi^{J}N^{\frac{J}{2}}g_{j}\leq\chi^{J}N^{\frac{J}{2}}
\]
 The theorem is proved.

\pagebreak


\begin{thebibliography}{10}
\bibitem{Golubev}Goloubev V. V. Lectures on analytic theory of differential
equations. Moscow, 1950 (in Russian).

\bibitem{Lanford} Lanford \textcyr{\char206}. Time evolution of large
classical systems. Lecture Notes in Physics, \textbf{38} (1975), 1-111.
Springer.

\bibitem{CerIllPul}C. Cercignani, R. Illner, M. Pulvirenti. The mathematical
theory of dilute gases. Springer. 1994. 

\bibitem{CaprPul}S. Caprino, M. Pulvirenti. A cluster expansion approach
to a one-dimensional Boltzman equation: a validity result. Comm. Math.
Phys., 1995, 166, 603-631.

\bibitem{DobrFri}R. Dobrushin, J. Fritz. Nonequilibrium dynamis of
one-dimensional infinite particle systems with a hard core interaction.
Comm. Math. Phys., 1977, v. 55, 275-292. 

\bibitem{CaMaPul}E. Caglioti, C. Marchioro, M. Pulvirenti. Non-equilibrium
Dynamics of Three-Dimensional Infinite Particle System. Comm. Math.
Phys., 2000, v. 215, No. 1, 25-43. 

\bibitem{Sin}Ya. Sinai. Construction of dynamics for one-dimensional
systems of statistical mechanics. Theor. and Mathem. Physics, 11,
1972, No. 2, 248-258. 

\bibitem{Spohn}H. Spohn. Large scale dynamics of interacting particles.
1991. Springer. 

\bibitem{mal_dynamicalClusters}V. \textcyr{\char192}. Malyshev. Dynamical
Clusters of Infinite Particle Dynamics. J. Math. Physics, 2005, v.
46, No. 7.

\bibitem{why_1}Malyshev V. A. Why the current flows: multi-particle
one-dimensional model. Theoretical and mathematical physics, 2008,
v. 155, No. 2, pp. 301-311. 

\bibitem{fixed_int}V. A. Malyshev. Fixed points for one-dimensional
particle systems with strong interaction. Moscow Math. Journal, 2012,
v. 12, No.1. 

\bibitem{fixed_per} Malyshev V. A. Critical states of multi-particle
systems with strong interaction on the circle. Problems of information
transmission, 47:2 (2011), 117-127. \end{thebibliography}
\end{document}